# Interfacial Effects during Phase Change in Multiple Levitated Tetrahydrofuran Hydrate Droplets

*Adam McElligott, André Guerra, Alexia Denoncourt, Alejandro D. Rey, Phillip Servio\**

Department of Chemical Engineering, McGill University, Montreal, Quebec H3A 0C5, Canada

\*phillip.servio@mcgill.ca

ABSTRACT

Recent strategies developed to examine the nucleation of crystal structures like tetrahydrofuran (THF) hydrates without the effects of a solid interface have included acoustic levitation, where only a liquid-gas interface initially exists. However, the ability now exists to levitate and freeze multiple droplets simultaneously, which could reveal inter-droplet effects and provide further insight into interfacial nucleation phenomena. In this study, using direct digital and infrared imaging techniques, the freezing of up to three simultaneous THF hydrate droplets was investigated for the first time. Nucleation was initiated at the aqueous solution-air interface. Two pseudo-heterogeneous mechanisms created additional nucleation interfaces: one from cavitation effects entraining microbubbles and another from subvisible ice particles, also called hydrate nucleating particles (HNPs), impacting the droplet surface. For systems containing droplets in both the second and third positions, nucleation was statistically simultaneous between all droplets. This effect may have been caused by the high liquid-solid interfacial pressures that developed at nucleation, causing some cracking in the initial hydrate shell around the droplet and releasing additional HNPs (now of hydrate) into the air. During crystallization, the THF hydrate droplets developed a completely white opacity, termed optical clarity loss or OCL. It was suggested that high hydrate growth rates within the droplet resulted in the capture of tiny air bubbles within the solid phase. In turn, light refraction through many smaller bubbles resulted in the OCL. These bubbles created structural inhomogeneities, which may explain how the volumetric expansion of the droplets upon complete solidification was 23.6% compared with 7.4% in pure, stationary THF hydrate systems. Finally, the thermal gradient that developed between the top and bottom of the droplet during melting resulted in a surface tension gradient along the air-liquid interface. In turn, convective cells developed within the droplet, causing it to spin rapidly about the horizontal axis.

# 1. INTRODUCTION

Studying the nucleation of crystal structures from aqueous solutions through physical experiments can impose several challenges. For instance, nucleation can be stochastic and, to accurately determine the mechanism and probability of formation, the measurement of a substantial number of individual local formation events is often necessary.[1] Moreover, multi-component systems that require gas or liquid diffusion for nucleation, in addition to energetic constraints, often have slower nucleation rates, and gathering representative data takes longer than in single-component liquid systems. Clathrate hydrates, a class of crystalline compounds that form when a gas or volatile liquid is trapped in a cage of water molecule hydrogen bonds, are often critical components of such systems.[2] Hydrates are increasingly being examined for novel industrial applications such as carbon dioxide sequestration and flue gas treatment.[2-4] However, they are most commonly investigated for use in the oil and gas industry, where natural gas hydrates could be used for energy transport and storage.[5, 6] Hydrate nucleation is akin to ice nucleation; it is divided into primary and secondary types. Primary nucleation occurs in systems that do not initially contain matter that has already crystallized. In contrast, secondary nucleation refers to systems where crystal nuclei form in the vicinity of pre-existing crystals.[7] Primary nucleation can be further subdivided into homogeneous and heterogeneous nucleation, mainly distinguished by their nucleation site. Homogeneous nucleation occurs in the liquid bulk, while heterogeneous nuclei form at structural inhomogeneities (i.e., surfaces and interfaces such as container walls, grain boundaries, or insoluble impurities).[8] Heterogeneous nucleation occurs more favourably at higher temperatures than homogeneous nucleation, so it is the most probable mechanism and dominant mode for nucleation.[8] However, this implies that system-specific defects on internal reactor surfaces can play a significant role in defining the rate of nucleation and hydrate growth.[9]

In turn, the effects of solid interfaces on hydrate formation measurements require additional consideration so that results can be generalized and novel hydrate systems can be scaled from the laboratory to the industrial level.

Previous experimental works which aimed to understand these solid-interface-specific effects have attempted to limit or eliminate the presence of that interface. In this way, they could examine how hydrate formation was modified through the absence of a solid substrate. For example, hydrates have been formed on stationary hydrophobic surfaces or suspended from the tips of fine glass filaments.[10, 11] Equally, hydrate growth was investigated in hydrophobic, hydrocarbon-oil-based suspensions.[12] Most notably, Jeong et al. (2022) formed natural gas hydrates with water droplets suspended in an acoustic levitator (i.e., levitated hydrates), where no solid interface was in contact with the liquid sample.[13] Their investigations determined that levitated hydrates had more nucleation sites than those grown from solid surfaces, though the nucleation work was greater.[9] Recent advances in acoustic levitation systems have allowed multiple liquid samples to be frozen simultaneously.[14] The TinyLev, developed by Marzo et al. (2017), is a single-axis, non-resonant levitator that can simultaneously levitate several droplets using multiple ultrasonic transducers.[15] By using a custom cryogun rather than a cooling chamber for freezing, the configuration of the levitator can remain open with no obstructing interface and allow for direct and clear image capture from digital or infrared (IR) cameras. Moreover, droplet size and location (called nodes) can be controlled.[14] Therefore, this system is ideal for further investigation into levitated nucleation and crystallization in hydrate systems free of solid interfaces.

This study will examine the morphological and thermal behaviour of water droplets containing 19.2 wt% tetrahydrofuran (THF), a stoichiometric ratio for hydrate formation equivalent to 5.6 mol%, suspended in an acoustic field during the phase change from aqueous solution to solid clathrate hydrate.[16] Up to three droplets will be frozen simultaneously under the direct observation of two synchronous cameras to determine any inter-droplet effects at the nucleation interface. To the best of our knowledge, this is the first time that THF hydrate formation and the effects of multiple concurrent hydrate-formation events have been examined in a levitation device. Unobstructed measurements of this type of system with coincident cameras are also entirely novel. Furthermore, this study will go beyond the focus of many previous investigations and examine levitated hydrate morphology after the nucleation and initial freezing stages: it will additionally look at bulk crystal growth (solidification) and melting effects. Therefore, dominant nucleation factors, nucleation mechanisms, and the interfacial and bulk transport phenomena present during crystallization and melting will be explored. In addition, this study will focus on the system's complex interactions between geometry, phase transition, and capillary processes at the primary interfaces (air-liquid, air-hydrate, and liquid-hydrate). Note that the open configuration does not allow for system pressurization. Therefore, THF hydrates, which form at atmospheric pressure and 4.4 °C, are present here rather than natural gas hydrates.[2] However, previous computational studies have shown that THF hydrates are suitable substitutes for natural gas hydrates as they share mechanical and vibrational properties: their structure-property relationships primarily rely on hydrogen bond properties rather than the guest molecule.[17] For instance, Kida et al. (2021) found that THF hydrates exhibited similar elastic moduli and uniaxial compressive strengths compared to natural gas hydrates.[18] The use of THF also differentiates this study from previous ones, as the guest molecule is not the primary component of the gas phase.

## 2. MATERIALS AND METHODS

### 2.1 Experimental Setup

This section describes the experimental levitation setup, which was first developed by Marzo et al. (2017) and modified by McElligott et al. (2022) for freezing.[14, 15] It is presented in **Figure 1** in a simplified form; further information can be found in the previously-mentioned sources. The levitation device is the TinyLev (E in the figure): two arrays spaced 7.5 cm apart consisting of 36 acoustic transducers arranged in rings. Each array has a hole in its center aligned vertically with the levitation axis. This configuration has eleven nodes for liquid droplet placement, and previous studies have frozen five droplets simultaneously while operating at 40 kHz.[14] However, only the three droplets in the center nodes will be examined in this study to improve their resolution during phase change (i.e., the cameras can be closer to the droplets, and the images will be clearer for analysis). The droplet positions are numbered 1 to 3 from top to bottom. A Delta Elektronika power supply (F, SM 70-AR-24) operating at 10 V and 0.8 A powers the TinyLev, creating the same acoustic field as in previous studies using pure water, which will be used for comparison.[14] The generated power runs through a driving board (G) which produces and amplifies the square wave excitation signals sent to each array. The average room temperature and humidity were 22.6 ± 0.6 °C and 23.8 ± 5.9 %, respectively. As the air supply to the system came from the room, this humidity was maintained through each experiment.

A bespoke cylindrical, stainless steel 316 cryogun (D) placed directly above the top array and centred on the axis of levitation is used to freeze the droplets. Positions 1, 2, and 3 are 6.4, 6.9, and 7.4 cm from the base of the cylinder, respectively. After the cylinder is filled with liquid nitrogen, the temperature at its base stabilizes to -55.5 °C. The air underneath it is then cooled such that it increases in density and begins to fall due to gravity. A significant cooling stream, maintained throughout each run, is generated from this process and results in liquid sample freezing. A Canon EOS 60D DSLR camera (18.0-megapixel CMOS sensor) equipped with an MP-E 65 mm f/2.8 1-5x macro lens and mounted onto an OptoSigma multi-axis manual translation stage (C) is used to take digital images of the droplets. Visualization of the droplets was optimized using a black backing inserted into the TinyLev and a fibre optic LED (A) for illumination. A Jenoptik IR-TCM 384 infrared camera calibrated for a 1–2% measuring accuracy in the range of -20 to 20 °C and with a NETD temperature resolution of less than 0.08 °C (B) is used to capture thermal images. This is how temperature was measured during each experimental run. Both cameras have lines of sight that are normal to the axis of levitation (i.e., parallel to the table) and perpendicular to each other, with the digital camera pointed at the black backing. Note that an HSI Fastcam Mini AX50 high-speed camera (not depicted) was used later to expand upon specific results gained throughout the experimental schedule.

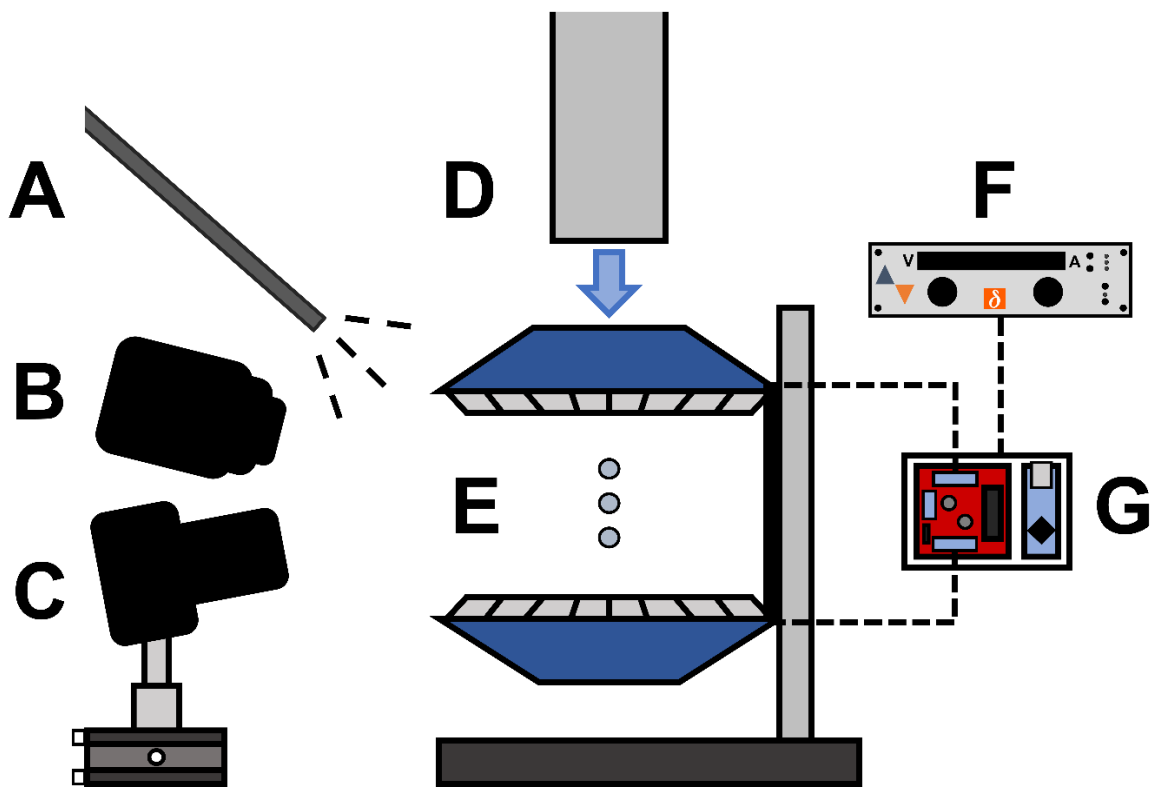

**Figure 1.** Simplified experimental setup schematic. The schematic includes an LED light source (A), infrared camera (B), digital camera with lens and multi-axis positioner (C), bespoke cryogun producing a cooling stream (D), TinyLev acoustic levitator with black backing (E), power supply (F), and driving board (G). During operation, the lines of sight of the digital and IR cameras are perpendicular and level with the droplets; their orientation in this figure is only for clarity.

2.2 Experimental Procedure

Before data acquisition, the power supply was turned on to initiate the acoustic field in the TinyLev. A 3 mL syringe was used to place the droplets in the three nodal positions. In addition to having individual droplets in the field, there were four multi-droplet configurations: positions 1 and 2 (1/2), 1 and 3 (1/3), 2 and 3 (2/3), or 1, 2, and 3 (1/2/3). The THF used in this study was obtained from Sigma-Aldrich (anhydrous, purity of 99.9%), and the remaining 80.8% of the droplet was reverse osmosis (RO) water. The first step in an experimental run was to fill the cryogun with liquid nitrogen. After one minute, to ensure adequate cooling of the cryogun surfaces, data capture from the digital and IR cameras was initiated, and the cryogun was placed above the opening of the TinyLev's top array. The cryogun created a sufficient cold air stream to freeze any present liquid sample(s). The cooling period was set to 90 seconds to ensure the entire droplet was frozen. After this period, the cryogun was removed, and the droplets were allowed to melt for one minute. The previous study using pure water had a cooling period of 180 seconds.[14] However, as the strength of the cold air stream is the same in this study, and THF hydrates form at approximately 4.4 °C at atmospheric pressure compared to 0 °C for water, the driving force for formation is much greater and 90 seconds is enough for complete freezing and examinations of post-solidification events.[17] A total of 15 replicates were performed for each nodal position and configuration. Note that melting was only examined for position 2 droplets, as previous studies have suggested that position and configuration in an acoustic field do not affect melting dynamics.[14] Therefore, 105 replicates were performed and analyzed for this study, of which 15 included melting.

## 3. RESULTS AND DISCUSSION

### 3.1 Nucleation of Levitated THF Hydrates

Droplets of a 19.2 wt% THF aqueous solution were suspended in an acoustic field and frozen by a cryogun to form clathrate hydrates. Systems containing one droplet, two droplets in a chosen configuration, or three droplets in all positions were examined. The first phases of freezing are shown in **Figure 2** for a 1/2/3 system in terms of morphological and thermal behaviour. The behaviours exhibited in the figure are characteristic of liquid droplets frozen in an acoustic field. All droplets displayed these characteristics in some form, regardless of their number or position.[14]

At 0 seconds, just before the placement of the cryogun, the droplets are at rest in the acoustic field and have an oblate spheroid shape. When the cryogun is placed approximately one second later, droplet eccentricity is reduced, and they circularize. Note that when the cryogun is added, some turbulence is induced in the system and digital image capture at 5.3 FPS has minor blurring effects. At this point in the thermal images, the backdrop behind the droplets becomes a darker shade of pink, which indicates the presence of a colder cooling stream. The droplets also adopt different colours in the IR images, indicating they are cooling. At around five seconds, hydrate nucleation occurs, though it does not necessarily occur simultaneously in all droplets. Nucleation is evident by the hydrate "shell," which develops around the droplets in the digital images. As hydrate nucleation is exothermic, this event is also accompanied by a rise in droplet temperature, evident in the IR portion of the figure.

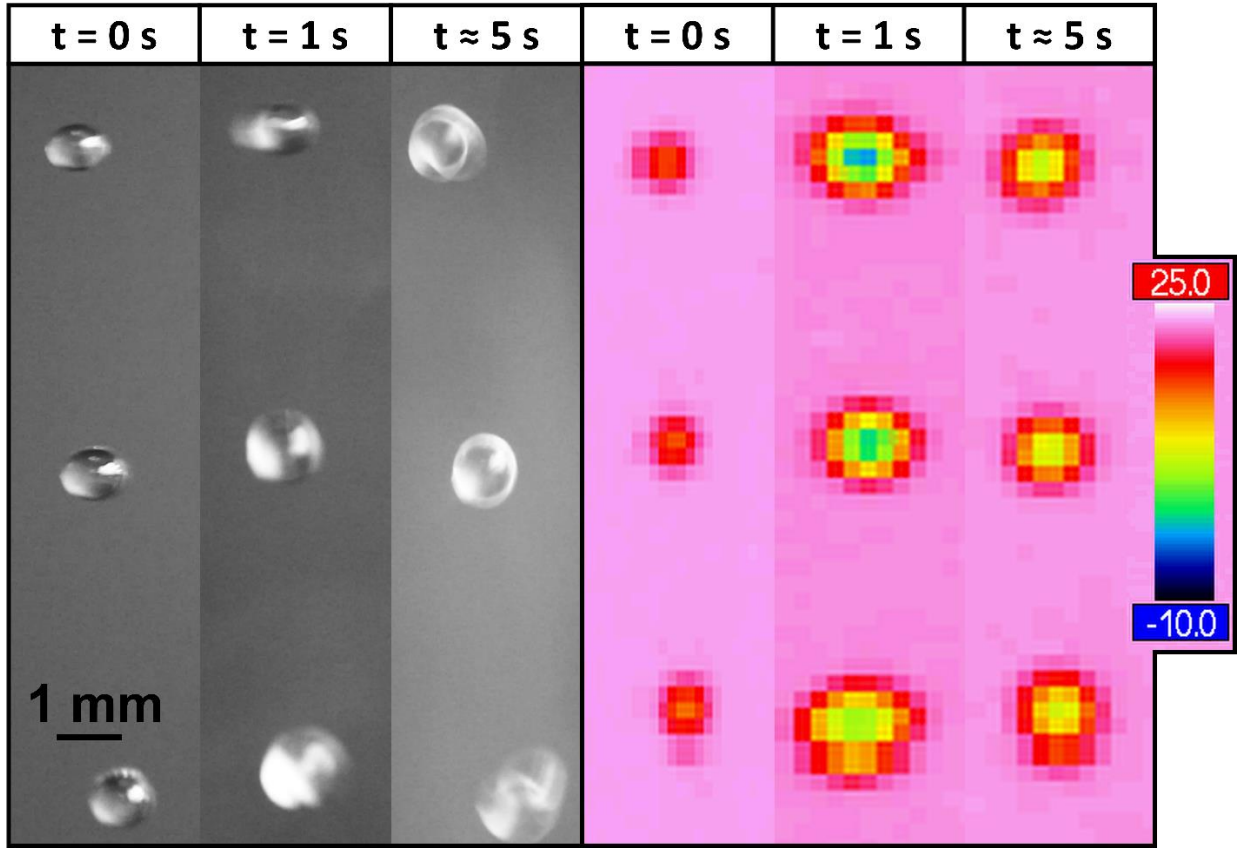

**Figure 2.** Morphological and thermal evolution from initial conditions to nucleation of THF hydrates in the 1/2/3 system. At 0 seconds, there is no cooling stream. At 1 second, the cryogun is added, and the droplets cool. Around 5 seconds, all droplets have nucleated.

3.1.1 Pre-Nucleation Conditions

While at rest in their nodal position, the droplets are not perfect spheres in the acoustic field. Instead, they adopt an oblate spheroid shape with a smaller vertical axis than the lateral one. This characteristic shape is caused by several forces present in the field. These primarily include gravity (downward), a counteracting acoustic pressure from the lower TinyLev array to maintain droplet stability, and acoustic pressure from the upper array that is used to prevent the droplets from leaving the acoustic region.[15, 19, 20] The combination of these forces creates acoustic streamlines at the boundary layer just before the droplet surface (the air-liquid interface) which determine the droplet boundary by creating vortices on its upper and lower halves.[14, 21] The droplet's shape is caused by the balance of these outer normal forces and the surface tension and hydrostatic forces within the droplet.[22] Acoustic streaming likely caused the solution to flow steadily at the surface, resulting in a small amount of evaporation and convective cooling.[23] This meant that the initial droplet temperatures were slightly lower than the average room temperature (22.6 °C): 21.5 ± 1.1 °C at position 1, 21.3 ± 0.7 °C at position 2 and 18.7 ± 0.7 °C at position 3. Previous pure water studies have observed about 1 °C variations between the droplet surface and environmental temperatures.[24] However, this does not explain the greater deviation at position 3. The droplets in position 1 and position 2 have similar average initial volumes of 0.63 ± 0.09 μL and 0.60 ± 0.07 μL, respectively. This value at position 3 is much larger, 0.97 ± 0.15 μL. The acoustic force of the lower TinyLev array is necessarily greater than that of the top as it must counteract gravity.[15] Therefore, it provides a stronger force on the closest droplet (position 3), which may allow it to have a higher volume without experiencing fragmentation. In turn, the larger surface area of the droplet would allow for greater acoustic streaming and so more evaporation at the surface.[14] This could lead to a further reduced initial temperature. Moreover, THF is a more volatile substance than water, which may have led to additional evaporation compared to droplets consisting purely of water. However, this effect was not observed in this study and is not likely present as this additional effect should have equally been seen for the position 1 and 2 droplets, which still contain THF but exhibit the same expected deviations as pure water systems.[24] It can be noted that compared to pure water in the same acoustic field, these droplets are around 1 μL smaller. This is because the aqueous THF-water solution has a lower surface tension.[25] Therefore, in an acoustic field with the same power supplied, the THF-water droplets must be smaller to remain stable (i.e., the same acoustic force applied to a droplet of similar volume but lower surface tension would cause that droplet to burst).

### 3.1.2 Individual Droplet Nucleation

When the cooling stream was added to the system, regardless of their number, each droplet experienced partial circularization from the initial oblate spheroid shape, which was maintained as their temperatures decreased. This reduction in eccentricity likely occurred as the stream added a downward lateral force, pushing it further against the lower array's strong acoustic pressure and altering the droplet's net forces. Nucleation was marked visually by two distinct and simultaneous events seen in **Figure 2**: the formation of a dendritic ice shell across the air-liquid interface and an increase in the temperature within the droplet and at its surface. Both phenomena are well-documented in previous studies on levitated crystallization, and hydrate nucleation is known to be exothermic.[2, 26-28] It is important to note that the airflow in this system rotates the droplet and results in a more uniform cooling at the air-liquid interface compared to most other levitated droplets, which are often stagnant in the acoustic field.[28] This means that the primary clathrate hydrate shell does not contain significant weak spots or bulges. The average droplet temperatures at nucleation were similar between position 1 and position 2, 4.38 ± 0.35 °C, while the same temperature at position 3 was 3.50 ± 0.99 °C. These correspond to similar decreases from the initial temperatures (recall that the initial temperature at position 3 was also lower), which indicates similar cooling rates were present at all three positions, about 3 °C per second. Note that these temperatures are too warm for ice nucleation on the droplet, which suggests that the droplets were mostly, if not entirely, made of THF hydrate when frozen. While sufficient temperatures for ice nucleation would eventually be reached, the majority of the droplet was frozen by that point.

Additionally, these temperatures are very close to the hydrate formation temperature, which indicates that a significant level of undercooling was not necessary for nucleation and that the energy barrier to hydrate nucleation may have been reduced. Therefore, it is useful to examine the nucleation time data as they may elucidate the nucleation mechanism.

**Table 1** shows the time required for nucleation for each droplet for all configurations now using the placement of the cryogun as the starting point. Individually, position 3 droplets, though furthest from the cryogun, nucleated fastest. These were followed by the position 1 droplets closest to the cryogun. Note that a statistical test (Welch's T-test) determined no significant difference between the two means at these positions. In other words, position 1 and position 3 droplets can be considered to have required the same amount of time to nucleate. This likely indicates that the primary shell growth between the two was also similar in crystallization rate. The position 2 droplets required a statistically significant, longer period before nucleation on average compared to the other positions, yet they had a similar nucleation temperature to position 1. This means that they may have had thinner, less opaque primary hydrate shells compared to the other droplets, though this could not be confirmed visually from the digital images obtained for this study. In all cases, the nucleation times are shorter than those for pure water droplets in the same system. This is logical as all system parameters (e.g., the cooling rate) were the same in both cases, but the driving force for nucleation is higher in the THF-water system as crystallization occurs at warmer temperatures.

**Table 1.** Nucleation time for droplets at each position in each configuration in seconds. The values in brackets are ($\pm$) the 95% confidence intervals.

|  | Individual | Positions 1/2 | Positions 1/3 | Positions 2/3 | Positions 1/2/3 |
|---|---|---|---|---|---|
| **Position 1** | 5.13 (0.78) | 4.79 (0.70) | 4.40 (0.53) | - | 4.73 (0.91) |
| **Position 2** | 6.80 (0.95) | 6.00 (0.82) | - | 4.50 (0.90) | 4.93 (0.94) |
| **Position 3** | 4.67 (0.83) | - | 4.53 (0.78) | 4.73 (0.72) | 5.00 (0.86) |

These individual nucleation time results could indicate the likely nucleation mechanisms. Homogeneous nucleation depends on the level of undercooling, which we can relate to the temperature at nucleation, rather than the droplet volume.[29] In other words, two droplets with different volumes at the same temperature can exhibit similar nucleation times. It follows that the position 3 droplet, having achieved the coldest nucleation temperature, nucleates first if the nucleation is homogeneous. However, it would also mean that the position 1 and 2 droplets should nucleate at similar times and after the position 3 droplet. Instead, there was no statistical difference between the position 1 and position 3 nucleation times, while nucleation at position 2 required more time. Therefore, homogeneous nucleation is likely not a significant nucleation mechanism. Instead, heterogeneous nucleation may be dominant in the system. The nucleation of ice in levitated droplets has previously been estimated to have a 90% probability of occurring on the surface.[23, 29] There could be a pseudo-heterogeneous mechanism at the air-liquid interface where ultrasonic waves create a concentrated acoustic pressure and cause cavitation. In turn, this promotes heterogeneous nucleation by increasing the number of microbubbles entrained at the surface.[23, 30, 31] Nucleation from the microbubbles at the air-liquid interface requires less work and has been reported as the most thermodynamically favourable and likely mechanism for the levitated nucleation of ice and natural gas hydrates.[9, 32] This nucleation phenomenon would occur at a scale too small to measure in the current study's system. However, the previous levitation studies that were able to provide direct evidence of the presence of this mechanism used similar ultrasonic frequencies (e.g., 34.5 or 39 kHz), indirectly indicating the likelihood of the microbubble mechanism in the present system.[9, 31] In this case, droplet surface area becomes the most influential parameter for nucleation, which can explain why the droplet with the largest volume (position 3) nucleates first. However, while this mechanism may be present in the system, it does not explain why the position 1 droplet, which has a smaller surface area, nucleates at similar times to position 3.

An additional pseudo-heterogeneous mechanism may have been present in the system that caused position 1 droplets to nucleate faster. Previous studies have suggested that nucleation in levitated ice in a relatively high humidity (> 20%) atmosphere, such as in our system, is largely caused by aerosolized, subvisible ice particles.[14, 29] These particles are formed from supercooled water in the air and are directed towards pressure nodes within levitation systems.[33] These hydrate-nucleating particles (HNPs) made of ice may also be causing nucleation in the system by forming and impacting the cooling liquid droplets. The presence of ice in the system may seem to contradict the reported hydrate nucleation temperatures, which are above zero. However, these subvisible particles have sufficiently small volumes to reach sub-zero temperatures before the significantly larger THF-water droplets, under the same cooling stream, reach sub-4.4 °C conditions.

Previous studies have demonstrated how small ice particles can cause hydrate nucleation.[34-36] Note that this is referred to as pseudo-heterogeneous nucleation and not secondary, as the (hydrate) crystal structure that forms in the droplet is different from the (ice) crystal structure present prior to nucleation (i.e., if ice were nucleating ice, it would be called secondary). When ice impacts the local water film, it acts as a template for hydrate structure and provides energy for hydrate nucleation.[34] At the ice-water interface, hydrogen bonding between ice and liquid water could lower the potential energy of interfacial water molecules that are part of the local solution phase, causing a transition layer of "mediator cages" or cage-like structural fluctuations to develop.[35, 36] The transition layer has a lower surface free energy than the ice-liquid interface, promoting the formation of amorphous nuclei that grow and evolve into complete THF hydrate cages.[35-37] This initial clathrate coating at the new ice-THF hydrate interface then grows perpendicular to that interface into the droplet.[35] Note that this does not mean that a significant portion of the droplet (i.e., what is visible in the nucleation frames of **Figure 2**) is made of ice. The length scale of the transition layer is in angstroms, and the timescale for the transition to occur is in the nanosecond range, compared to the millimetre and millisecond scales measured in our study.[35] Moreover, formation occurs above 0 °C and only at driving forces for THF hydrate formation. Therefore, it is unlikely that any significant portion of the droplets is made of ice, even if the impact of an ice structure causes nucleation.

In the case of our system, nucleation from HNPs is a function of air-liquid interfacial area (for impact) and distance from the cryogun, as there may be more HNPs closer to the cryogun due to either frost or colder atmospheric temperatures. This could explain how droplets with a higher surface area (position 3) and nearest to the cryogun (position 1) can nucleate simultaneously: nucleation via HNPs is promoted in both cases. However, it is difficult to draw strong conclusions about the dominant pseudo-heterogeneous nucleation mechanism at all positions. Assuming that position 3 nucleates first because it has the largest surface area, it is unclear which mechanism is the cause of nucleation in most cases, as both types are promoted by additional cavitation or impact sites. Now adding that position 1 nucleates at the same time as position 3 despite being smaller, it is likely that the HNP mechanism is dominant because that droplet is closer to the cryogun. Finally, the droplet at position 2 nucleates last and is farther from the cryogun than position 1 and smaller than position 3. Again, as these changes would restrict both mechanisms, it is unclear from the data which one is predominant at that position. In short, while the results suggest nucleation in the system is likely pseudo-heterogeneous in all cases, only the HNP mechanism at position 1 is expected to be dominant. Computational modelling studies are recommended for future work: a dominant mechanism could be determined by comparing the local nucleation energy provided by cavitation with that provided from HNP impact. Note that, as stated in the Introduction, it is often posited that levitation eliminates system-specific experimental nucleation effects from, for instance, microscopic defects in reactor walls. These results suggest that nucleation is affected by system-specific factors like acoustic field strength and humidity. However, unlike microscopic wall defects, it is relatively simple and easy to quantify these factors. Therefore, it is more accurate to say that levitated crystallization replaces system-specific effects in reactor/crystallizer systems that are difficult to quantify with more straightforward and measurable ones.

3.1.3 Multiple Droplet Nucleation

Turning towards the other columns in **Table 1**, the multi-drop configurations, there is a notable change in the nucleation time at position 2. While these times at position 1 and position 3 (statistically) are constant regardless of configuration, the position 2 nucleation times decrease in the 2/3 and 1/2/3 systems. Furthermore, the nucleation times across positions in the 2/3 and 1/2/3 systems are statistically constant: all droplets nucleate simultaneously. Note that this is also the case in the 1/3 system, but as position 1 and position 3 droplets nucleate at similar times on an individual basis, this is not a notable change. Near-simultaneously droplet nucleation may be within the error of the digital camera framerate, so a high-speed camera at 2000 FPS was used on the 2/3 system to determine a precise difference in nucleation times. Examples of the resulting images are available in the Supporting Information. It was found that there were 419 frames on average between the nucleation at position 3 (first) and position 2, which is 0.21 (± 0.06) seconds. This is similar to the difference of the 2/3 system averages (**Table 1**) and smaller than the 2.13 s difference between the individual droplets. This result suggests that the coincident nucleation in the 2/3 and 1/2/3 systems is more than just a product of a lower framerate. Therefore, supplemental nucleation effects could be present in multi-drop systems in addition to the two pseudo-heterogeneous mechanisms previously described. The specific volume of THF hydrate is greater than that of water. Therefore, after the initial dendritic THF hydrate shell forms around the air-liquid interface, the shell grows inwardly and significantly increases interfacial pressure. This increase could impost a sufficiently great mechanical stress that results in shell deformation or cracking.[26, 38] The latter can cause the emission of additional HNPs (now made of THF hydrate rather than ice) to the atmosphere that are directed towards the nearest-neighbour droplets by the acoustic field. Moreover, the cooling stream enhances the likelihood of cracking by improving latent heat transport by forced convection at the air-hydrate interface and increasing the hydrate growth rate.[28] This would also increase the mechanical stress on the shell and produce a higher fragmentation frequency. In short, the post-nucleation emission of HNPs may induce nucleation at position 2 in multi-droplet configurations, resulting in shorter and statistically similar nucleation times. Consequently, it is more accurate to say that there is no new nucleation effect in multi-drop systems, but the HNP pseudo-heterogeneous mechanism is strengthened and becomes dominant at position 2. Note that this does not apply to the 1/3 system where the droplets are further apart (non-adjacent) and additional HNPs may be directed to the empty position 2 node.

However, it cannot be said that droplets in adjacent positions necessarily cause each other to nucleate, either, as this does not occur in the 1/2 system. This is different from pure water systems, which showed statistically similar nucleation times in all configurations with adjacent droplets.[14] The difference may arise from three factors. First, the averages obscure that one-fifth of the droplets (3/15) nucleated simultaneously, compared to 10/15 in water (again, in the 1/2 system).[14] As such, it is more precise to say that droplets in adjacent positions only have a higher likelihood of nucleating together. Second, this likelihood may be reduced for THF hydrates as they have greater tensile strengths than water (i.e., they have previously been measured to crack only under liquid-solid interfacial pressures in the gigapascal range versus the megapascal range for ice), which means that there could be less cracking and less HNP emission.[39-41] However, this implies that the instantaneous pressures at the liquid-solid interface at nucleation between ice and THF hydrate are similar, which may not necessarily be the case. While models exist to calculate these values for water, they cannot be applied to hydrates due to significant differences in growth rates and mechanisms, and no models for this hydrate formation parameter currently exist.[27] Nonetheless, considering the short timeframes and small length scales of this investigation, particularly as they pertain to hydrate shell cracking, it is unlikely that there are significant pressure differences compared to pure water, so it is probable that fewer HNPs are produced. Further evidence for this supposition is provided in the next section. Finally, in addition to the latter, position 1 droplets have a significantly smaller surface area than position 3 droplets, which could reduce the amount of post-nucleation HNPs produced compared to the 2/3 and 1/2/3 systems. Paired with a greater tensile strength, this would further reduce the likelihood of inter-droplet effects in the 1/2 system.

### 3.2 Solidification of Levitated THF Hydrates

The remaining crystallization stages, up to complete solidification, are presented in **Figure 3** for the 1/2 system in terms of morphological and thermal behaviour. Again, the behaviour depicted is characteristic of all droplets regardless of droplet number or configuration. After droplet nucleation (A in the figure), the hydrate shell grew inwardly towards the central liquid bulk. Soon after, the droplets would lose their optical clarity and become opaque white spheres (B). This optical clarity loss (OCL) also marked the beginning of a significant reduction in droplet temperature. Moreover, protrusions could begin to grow on the air-hydrate interface (C) as the droplet continued to solidify. It is crucial to note here that, despite the order of how they occur in the figure, protrusion formation did not necessarily begin after OCL. While the OCL always occurred, not all droplets formed protrusions. Additionally, no correlation could be determined between the times until nucleation, OCL, or the development of initial protrusions. Therefore, these latter two events likely occur through different phenomena. Finally, the droplet would completely crystallize (D), and no liquid solution likely remained as the cold temperature was maintained (i.e., there was no further heat generation from stoichiometric hydrate growth).

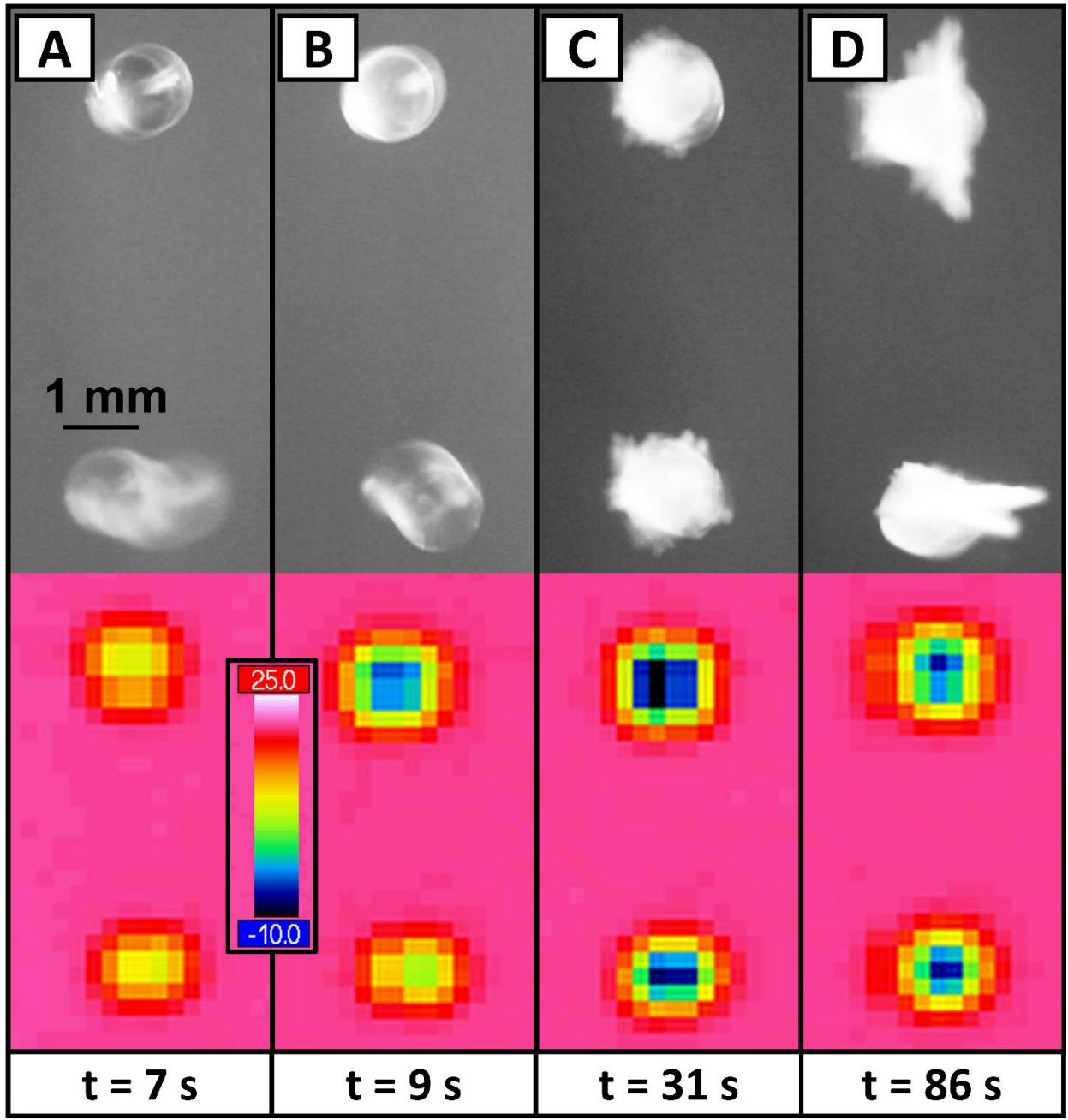

**Figure 3.** Morphological and thermal evolution from nucleation to solidification of THF hydrates in the 1/2 system. After the droplets have nucleated (A), significant inner hydrate growth causes a loss of optical clarity in position 1 (B). Eventually, both droplets lose clarity and can form protrusions (C), which grow as the droplets freeze completely (D).

### 3.2.1 Optical Clarity Loss

A few seconds after nucleation, the originally clear droplet would develop a white inner opacity that endured for the remainder of the freezing time. Initially, it was supposed that this optical clarity loss was related to the levitated hydrate growth rate: after sufficient time, the THF hydrate shell would have grown thick enough to modify the overall optical properties of the droplet and modify light's path as it crossed the air-hydrate and then hydrate-liquid interfaces. This notion was bolstered by the cooling effect that also occurred immediately upon OCL, which is evident in **Figure 3**B, where there was a significant temperature reduction compared to the post-nucleation temperature in **Figure 3**A. This indicated that most of the liquid bulk had been changed to the hydrate phase and that the constant cooling rate exceeded the exothermic heat release from the remaining crystal growth. However, while a correlation with the growth rate is likely, the origin of the OCL remains unclear. Bulk freezing in levitated pure water droplets resulted in clear ice formation in similar or longer timeframes.[14] Moreover, the refractive index of gas hydrates is very similar, 1.35 compared to 1.31 for ice, indicating that clear frozen droplets should still be expected here.[42] The OCL may be related to trapped bubbles in the crystal matrix. The solubility of air in a crystal phase (e.g., ice or hydrate) is several orders of magnitude smaller than in liquid water.[43] Therefore, as the solid front grows towards the droplet center, the concentration of dissolved air in the liquid phase (at the solid-liquid interface) becomes enriched until bubbles form.[44] If the growth of the solid front is relatively slow, it can push all the air into a single bubble or a small number of large bubbles in the droplet center, which is what occurred in the pure water system.[14] However, a much faster crystal growth rate would result in many smaller bubbles becoming trapped within the solid phase, significantly reducing droplet clarity by creating several microscale solid-air interfaces that repeatedly refract light. As mentioned, the THF hydrate growth rate in this study is greater than the ice formation rate in previous pure water studies because the driving force for formation is higher and, again, ice formed in similar conditions did not exhibit OCL.[14] Therefore, the OCL mechanism may develop from a specific hydrate shell thickness that is reached at a sufficiently quick rate to produce inhomogeneities in the crystal structure. Note that, while the OCL can be used as a measure of the hydrate growth rate, it is a limited one as it requires high bubble entrapment rates to produce a visual effect. Examining the rate through thermal effects (i.e., the start of the cooling period where the heat from hydrate formation is reduced) is likely a more accurate method with a broader applicability.

THF hydrate growth rates varied at the different positions, though the number of droplets did not have an effect. Quantifying these rates through the OCL is possible in this study, as all droplets exhibited this visual phenomenon. Specifically, the nucleation-adjusted OCL times (i.e., the OCL time with the nucleation time subtracted to remove variation between positions) were $6.7 \pm 1.4$ s, $9.5 \pm 2.0$ s, and $11.0 \pm 1.8$ s for the position 1, position 2, and position 3 droplets, respectively. These values may provide some insight into the effect of the nucleation mechanism on the hydrate growth rate. It was previously suggested that ice HNPs were the dominant pseudo-heterogeneous nucleation mechanism at position 1, which also has the fastest OCL. Moreover, HNPs likely caused nucleation at position 2 in the 2/3 and 1/2/3 systems, which has the second-fastest (regardless of system) time. The HNP mechanism may induce faster growth rates as ice provides an initial structural template at the ice-liquid interface. In contrast, cavitation-induced nucleation requires additional energy for molecular rearrangement at the air-liquid interface.[34, 36] However, it should be noted that this correlation is weak as the droplets and systems mentioned only account for half of the total number of droplets investigated. The correlation does not explain why position 2 droplets, individually or in the 1/2 system, have similar growth rates to their counterparts in other systems. In addition, it is possible that position 3, where the dominant nucleation mechanism is unclear, has the slowest growth rate because it has the greatest liquid bulk and warmer inner droplet temperatures that reduce the growth rate. In other words, interfacial nucleation effects on the growth rate may be limited compared to other factors such as volume or proximity to the cryogun. Therefore, while there is some evidence that HNP nucleation results in faster hydrate growth rates at position 1 and position 2, there is insufficient data to suggest that the nucleation mechanism is the most influential component of the growth rate.

### 3.2.2 Interfacial Protrusion Growth

In addition to bulk growth, it was possible for several protrusions, like in **Figure 3**C and continuing in **Figure 3**D, to form at the hydrate-air interface. Though OCL effects always occurred, protrusion formation and growth appeared stochastic: droplet position did not affect the frequency of protrusion formation, and there was no trend in the time between nucleation and initial protrusion formation (i.e., protrusions could form anywhere from 8 to 35 seconds after nucleation under the same conditions, a tremendous time frame in the context of this study). In other words, the protrusion growth shown in the figure is demonstrative but not indicative of all cases: only about one-third of the total droplets tested produced protrusions. Moreover, due to the stochasticity at the individual positions and low formation frequency, it could not be determined whether the presence of multiple droplets affected the protrusion frequency. Even in the multi-drop systems, protrusion formation did not always occur, and it was possible that they only formed on one of the two or three droplets present. This is opposed to the pure water system, where protrusions always formed, were larger, and occurred more quickly when multiple droplets were present.[14, 45] Several previous studies using pure water have suggested that protrusions are seeded by ice either from the air or produced from interfacial cracking at nucleation (akin to HNPs in this study).[14, 26] Heat evolution from crystal growth may produce a quasi-liquid layer at the hydrate-air interface from which protrusions could grow.[46] Therefore, in the case of the current work, the protrusions could be an agglomeration of small ice and hydrate particles directed toward (or back toward) the droplet surface by the acoustic field. However, the lower frequency of protrusion formation could indicate fewer HNPs than in the pure water system. As the same amount of atmospheric water, and so ice particles, are present in both studies, the lower volume of crystals in the air would suggest that there are fewer post-nucleation HNPs (the only other HNP source) from less cracking at nucleation. This is evidence that the internal pressure at nucleation is relatively similar between ice and THF hydrates, as the higher hydrate tensile strength would result in less cracking and less HNP production for a similar pressure. Therefore, there would be a lower likelihood of (1) protrusion formation and (2) neighbouring droplets affecting each other's nucleation times, both of which were observed. Moreover, Jeong et al. (2019), the only other group to have performed levitated hydrate experiments to this point, suggested that protrusions on levitated natural gas hydrates form from the bulk liquid.[13] Briefly, the internal droplet pressure may push the interior liquid out of the droplet through defects or pores in the hydrate shell, driving discrete localized growth events: a mechanism which has also been seen in previous pure water studies. A similar protrusion formation mechanism is also possible in the current system, as THF and natural gas hydrate mechanical properties are similar, though driving forces are lower here.[17] Again, however, the lower formation frequency may indicate that any pressure increases from hydrate nucleation to ice nucleation are small compared to the tensile strength increases from THF hydrate to ice. This would lead to fewer protrusions in the hydrate system, as similar pressures against superior hydrate mechanical properties would result in less liquid pushed through the solid shell. In short, the lower occurrence of protrusions may indicate that the stronger hydrate shell limits the additional production of HNPs and could explain why THF hydrate droplets have a reduced likelihood of affecting each other compared to ice droplets that have weaker shells.

### 3.2.3 Volumetric Expansion and Sphericity

When the 90-second experimental run had been completed, the droplets were completely frozen, and no liquid remained. These hydrate droplets either looked like what is present in **Figure 3**D if protrusions formed or simply as white spheres if they did not. It is notable that when examining the thermal images from **Figure 3**C to **Figure 3**D, there was a slight warming effect at the droplet surface and generally less of a thermal gradient across the droplet. This indicates that as the 90-second mark was approached, there was a slight diminishment in the cryogun cooling stream as the liquid nitrogen was used up. While this diminishment was likely present in the pure water system, no warming was observed.[14] This may relate to the decreased frequency and size of protrusions in the THF hydrate system. When there are more dense protrusions, they can act as an insulating layer around the droplet, allowing it to maintain colder temperatures. These were not present for the THF droplets, resulting in a small temperature increase.

During the phase change from aqueous solution to solid THF hydrate, a volumetric expansion of 7.4% was expected for a pure, stationary system.[47] In this study, the measured volumetric expansion significantly exceeded this value: 23.6 ± 1.3 % on average. This value was consistent regardless of position or number of droplets and is not affected by the larger volume at position 3 because it is a percentage increase. Furthermore, protrusions were not factored into the frozen volume; only the spherical "core" of the droplet was taken. This was because previous water studies also only looked at the "core" expansion for comparison, which is essentially an examination of the elongation of the hydrate-air interface, and it could not be determined if the protrusions consisted only of hydrate.[14] The volumetric expansion increase is likely the result of non-ideal environmental factors present in open levitated systems (i.e., droplet purity and acoustic pressure-induced oscillation). The air that becomes entrapped in the crystal lattice of the hydrate phase is included in the volume measurement and would result in a higher final volume than pure THF-water droplets, which do not contain air. Moreover, the kinetics at the inner boundary between the growing hydrate and liquid phases may be frequently modified by droplet oscillations in the acoustic field, particularly when the droplet is mostly still in the liquid state. This could also cause significant defects in the crystal structure, resulting in an effectively lower droplet density than the expected density of THF hydrates in a stationary medium and augmenting volumetric expansion.[7, 48] In a water-only system (ice growth), previous studies have measured volumetric expansions of up to 30.8% compared to the expected 9% for pure, stationary ice-forming systems, which is a significantly greater increase in expansion compared to what was observed in this study.[14] Similar air concentrations and oscillatory amplitudes at the phase boundary are present in open levitation systems regardless of which solution is crystallized, so changes in volumetric expansion likely come from differences in crystal structure and mechanical properties. In this case, the bubbles are more numerous but smaller due to faster crystal growth and may contribute less to the final volume. In addition, the stronger mechanical properties (higher tensile strength) of THF hydrates compared to ice may dampen oscillatory effects at the phase boundary, reducing the magnitude of structural defects produced by the acoustic field. These changes in growth kinetics and solid-phase properties likely result in less volumetric expansion compared to water alone.

Another morphological measurement for hydrate-air interfacial crystal growth is the final droplet sphericity. This value is the ratio between the surface areas of a volume-equivalent sphere and the measured particle (when freezing is complete).[49] Note again that protrusions do not factor into sphericity measurements. In this study, the final droplet sphericity was always measured to have a value of 1. In other words, the solid droplet "cores" remained spheres regardless of position or droplet number. This behaviour is evident in **Figure 3**D, where any apparent interfacial elongation is a framerate effect. This does not mean that the droplets were perfectly smooth spheres, but the measurement accuracy in this study is in fractions of millimetres, so the microscale defects (or smaller) that are likely present are considered negligible. This is compared to pure water, where defects were significant, measurable at scale, and resulted in reductions in sphericity: as low as 0.65 in the 1/2/3 system.[14] Several limited or non-existent factors in a hydrate system affected final ice droplet sphericity. The most important factor for ice droplet sphericity was the formation of a large bead at the bottom of the suspended droplet. This was caused by water escaping through pores in the ice shell and flowing downward due to gravity but remaining on the droplet due to surface tension before freezing. This behaviour was not observed to any significant extent in this study. The greater mechanical properties for similar pressures resulted in less, if any, of the aqueous phase escaping the droplet during solidification. In conjunction with pore effects, crystal growth rates are a critical factor for sphericity. Faster growth rates (i.e., THF hydrate compared to ice growth rates in the same system) can result in a solid hydrate shell that thickens more quickly, giving less chance for the liquid to escape the droplet. In addition, a more rapidly thickening shell limits the amount of liquid present during the experiment. As any acoustic oscillations affecting sphericity are more prominent in the liquid phase, this could reduce their impact. Finally, it has been suggested that protrusion formation can affect droplet sphericity if it forms early enough at the hydrate-air interface, even if not directly considered in the measurement. Previous studies have shown that with greater protrusion frequency and prominence, so greater

HNP presence, sphericity decreased.[14] It was proposed that the formation of protrusions on a sufficiently thin ice shell could modify further growth of the shell on both interfaces (with air or with the aqueous phase). However, as explained earlier, this system's production of HNPs was limited as less cracking was observed, and protrusions could often take significant time to form. Therefore, it is likely that the factors that cause sphericity reductions in ice are restricted in hydrate systems, so levitated hydrate droplets under a cooling stream remain predominantly spheres during the crystallization process.

3.3 Melting of Levitated THF Hydrates

After the crystallization period in the position 2 system, the cooling stream was removed, and the droplets were exposed to room temperature. They were then observed for one minute as they returned to the liquid state, as in **Figure 4**. Very quickly, the droplet became more oblate, losing its surface eccentricity and any protrusions (B in the figure). As the droplet continued to melt, it would eventually develop convection currents and spin about the lateral axis (C). In due course, only the liquid phase would remain (i.e., no THF hydrate was present), and the droplet returned to its initial temperature (D). However, despite returning to this temperature within the measurement period, droplet spinning persisted beyond one minute due to momentum.

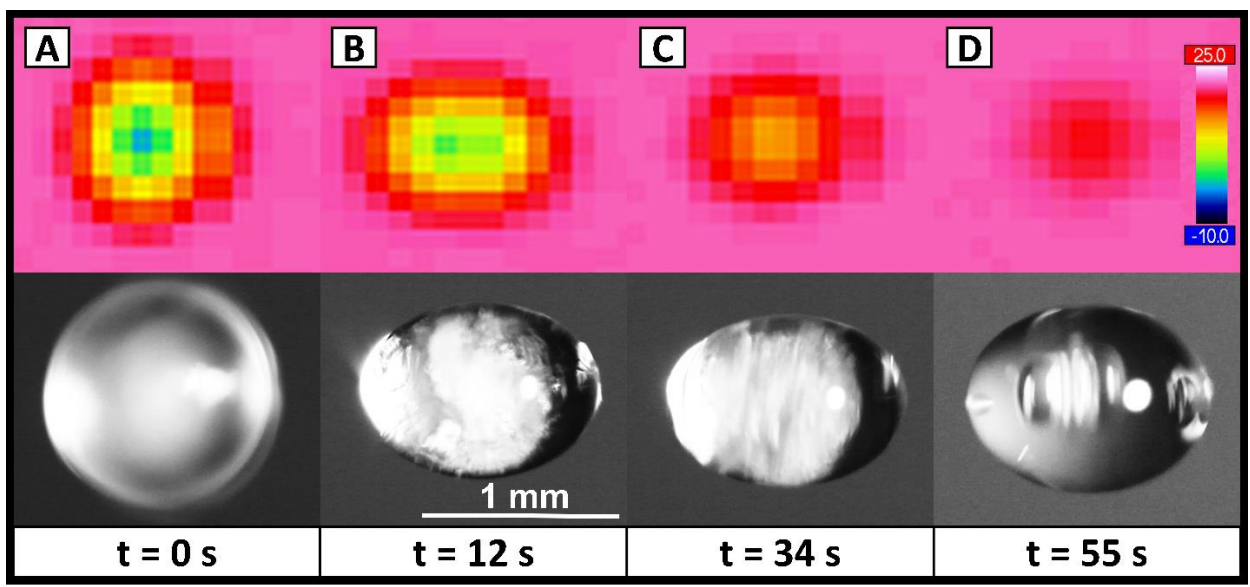

**Figure 4.** Morphological and thermal evolution from solid THF hydrate to a fully melted droplet in the position 2 system. First, the cooling stream is removed (A), and the droplet regains the oblate spheroid shape (B). Eventually, the droplet begins to spin (C) as it returns to its initial temperature (D).

When the cooling action was removed from the system, the frozen droplets were exposed to room temperature air, which resulted in an approximately 18 °C driving force for melting at the solid-gas interface (i.e., the difference between the room and thermodynamic melting temperatures was about 18 °C). Due to the rapid increase in temperature, the droplet surface and protrusions liquefied quickly. This resulted in a loss of spherical structure, and the droplet returned to a more oblate spheroid shape with a large portion of THF hydrate at its center. However, as seen in **Figure 4**B, the droplet was still somewhat angular due to the size of the remaining hydrate. After 20 to 30 seconds, when more of the central solid portion had melted, the remaining hydrate began to spin rapidly about the lateral axis, such as in **Figure 4**C. This behaviour is well-documented and common not only in levitated systems but in systems with melting at a gas-liquid interface generally.[14, 50] In brief, when the buoyant force brings the remaining solid phase to the top of the droplet, this creates a vertical thermal gradient where, in this case, the top is cooled by the THF hydrate, and the bottom is warmed by the gaseous interface. In turn, a sufficiently great surface tension gradient develops at the air-liquid interface such that convective currents, which originated at the droplet center and moved towards the bottom of the droplet, became convective cells that spun about the horizontal axis.[50] This form of surface tension-driven flow due to a thermal gradient is known as Bénard-Marangoni convection, and systems exhibiting this behaviour tend towards larger Marangoni numbers (Ma). Calculating this value as in previous pure water studies results in a Ma value of 6.9 x $10^3$, which is considered large enough to indicate that the convection cells result from the thermal gradient.[14, 51] Note that because the buoyant force always pushes the remaining hydrate upwards, the thermal gradient is always vertical and the axis of rotation for the spinning does not oscillate. Droplet spinning persisted after the solid phase had fully melted and beyond the measurement period. This was observed visually by the bubbles in the droplet continuing to spin despite a thermal gradient no longer being present in the droplet (**Figure 4**D). This was likely an effect where the intra-droplet currents, possibly assisted by acoustic streaming, had sufficient momentum to persist for a short time after the liquid temperature became homogeneous. Spinning beyond the melting period was not observed in the pure water study.[14] However, it was likely present in that study: the smaller and more numerous bubbles in the melted THF hydrate systems made spinning more apparent.

Continuing the comparison with the pure water system, the convection currents formed from ice melting had a reported Ma of 1.9 x $10^4$.[14] Therefore, the thermal gradient-driven convection cells in water likely had a greater velocity than in the aqueous THF system. There are two critical reasons for this. First, there is a greater driving force for melting in systems with lower freezing temperatures. In other words, when both solid phases are exposed to the same room temperature, there is more impetus for ice to melt as it freezes at 0 °C compared to 4.4 °C for THF hydrate. In turn, the initial temperature gradient during melting is more substantial in ice. Second, the thermal gradient results in a surface tension gradient that causes convective flow. On average, an aqueous THF solution will have a lower surface tension than pure water (as THF has a lower surface tension than water) and, therefore, weaker surface tension-driven flow. There were also possible convection-enhancing effects not accounted for in the Marangoni number. When protrusions formed, their interaction with the forces of the acoustic field and the cooling stream caused them to align with the horizontal axis, and the droplets would then spin about the (vertical) axis of levitation. It was observed that when the cooling stream was removed, protrusion alignment was shifted 90 degrees to the vertical axis, such that the acoustic field now aided droplet spinning about the horizontal axis. As protrusions were denser and more frequent in water freezing, they likely spun faster during melting than THF hydrate droplets, indicating that they probably continued to spin beyond the one-minute melting time. However, this does not mean protrusion-less THF hydrate droplets did not spin during freezing. Examining the digital images going from **Figure 4**A to **Figure 4**B, it is evident that there are clear portions of the droplet. Therefore, the OCL may have been a combination of the formation of many small bubbles (also visible in **Figure 4**B) and the THF hydrate droplet spinning due to the cooling stream. Consequently, the difference in this study is that when the cooling stream was removed, the THF hydrate droplet ceased to spin, so the droplet spinning during melting was initiated solely by the thermal gradient.

Droplet melting was considered complete when they returned to their initial temperatures, as in **Figure 4**D. The final volume of the droplets was measured to be 0.54 ± 0.09 μL on average, compared to the initial volume of 0.60 ± 0.07 μL. These values are not statistically different and indicate that a volume reduction during the freezing/melting cycle is unlikely. Changes in mass are possible in this study as volume can be added from protrusions (if formed from atmospheric water) and lost due to evaporation in the acoustic field (acoustic streaming). Mass losses could also result if protrusions formed from hydrate rather than water and were dislodged by the acoustic field. However, protrusion effects on the final volume are likely insignificant as very few protrusions of notable size were observed. Evaporation losses have previously been estimated to be 1.5% of the initial volume in pure water systems, and previous studies have found no change in the volume of water during a single levitated freeze/thaw cycle.[14, 30, 52] THF is more volatile than water, and evaporates more readily. However, the droplet is about 80% water, and even if the evaporation were doubled (an overestimate), the volume loss would remain within error at 0.02 µL. In other words, the evaporation of THF may be sufficient to cool the droplet surface before freezing but is not so significant as to change the droplet volume measurably. Therefore, it is possible that all these factors are present, and they add and subtract from the volume in amounts that result in it being statistically constant. However, they are all likely negligible: the liquid volume and component weight fractions did not change significantly from before nucleation to after melting.

## 4. CONCLUSIONS

The methods to eliminate container-driven effects on nucleation have recently been expanded to include acoustic levitation. For examining hydrate nucleation, previous investigations have been limited to single droplets contained within a cooling chamber.[9, 13] Expanding on these foundational studies, aqueous solutions containing 19.2 wt% THF were frozen to form THF hydrate while levitated in different configurations of three positions. Digital and thermal images were captured simultaneously of the single-, double-, or triple-droplet systems. This was the first time multiple levitated hydrate droplets were frozen, levitated THF hydrates were formed, and their crystallization was captured through unobstructed imaging. Melting of levitated hydrates was also examined for the first time. By examining the nucleation times of individual droplets, it was determined that nucleation occurred at the liquid-air interface via two potential pseudo-heterogeneous mechanisms. First, pressure from the acoustic field may have caused cavitation at the droplet surface, increasing the number of microbubbles from which the hydrate phase could nucleate. Second, subvisible ice particles from the atmosphere (called hydrate nucleating particles or HNPs) might have impacted the surface and initiated hydrate growth. It was not determined if any of these were dominant at all positions. Droplets in the 2/3 and 1/2/3 multi-droplet configurations had statistically similar nucleation times (i.e., they essentially nucleated together), indicating the presence of inter-droplet effects. The initial THF hydrate shell at position 3 likely experienced some cracking immediately after nucleation due to a rapid pressure increase at the liquid-solid interface. This released additional HNPs towards adjacent droplets that caused nucleation. However, this behaviour did not occur in the 1/2 system, which also had adjacent droplets, possibly because of the high tensile strength of THF hydrates, which limited cracking. As position 1 and 2 droplets had lower volumes than position 3 (reduced HNP production potential from cracking), inter-droplet effects from those positions were severely limited. This is compared to water, where inter-droplet nucleation effects were common in all multi-droplet configurations, as the proportionately smaller tensile strength of ice allowed for an increased incidence of cracking and even droplet breakup. The reduction of crystal particles in the atmosphere also reduced the frequency of protrusion formation in THF hydrates compared to ice, where frozen THF hydrate droplets tended to remain as relatively smooth spheres.

A few seconds after nucleation, the THF hydrate droplets would experience an optical clarity loss (OCL) and become opaque. This was ascribed to the high hydrate growth rate within the droplet, which did not allow time for air to diffuse further into the remaining liquid bulk, causing many small bubbles to be trapped by the growing solid front and act as structural inhomogeneities. Light refraction through all these small bubbles and droplet spinning from the cooling stream resulted in the OCL. Using the OCL as a growth rate measure indicated that nucleation resulting from HNPs may have been faster than that from cavitation, but the significance of the nucleation mechanism on OCL times could not be determined. When fully frozen, the THF hydrate droplets experienced a volumetric expansion of 23.6% compared to the expected 7.4%. This may have resulted from air bubbles or acoustic oscillations at the hydrate-liquid interface and could have been greater if not for the strong hydrate mechanical properties. Finally, during melting, the THF hydrate droplets would spin rapidly about the lateral axis, which was attributed to thermocapillary effects along the gas-liquid interface, creating intra-droplet convection cells. Momentum effects resulted in the droplets continuing to spin beyond the melting time and after returning to their initial temperatures. This study showed that inter-droplet effects exist between nucleating THF hydrate droplets and modify the liquid-to-hydrate phase transition. The insights presented here go beyond the crystallization of individual droplets and further our understanding of interfacial effects during hydrate nucleation, growth, and melting. Based on these results, future modelling studies could explore the pressure effects at the solid-hydrate interface upon nucleation, while future experimental studies could investigate how the presence of solid colloidal suspensions within the droplets modify interfacial crystallization.

AUTHOR INFORMATION

**Corresponding Author**

*phillip.servio@mcgill.ca

**Author Contributions**

The manuscript was written through the contributions of all authors. All authors have approved the final version of the manuscript.

ACKNOWLEDGEMENTS

The authors would like to acknowledge the financial support from the Natural Sciences and Engineering Research Council of Canada (NSERC) and the Faculty of Engineering of McGill University (MEDA, Vadasz Scholars Program). We also wish to thank Anne-Marie Kietzig and the Biomemetic Surface Engineering Lab at McGill for the use of their TinyLev device.

SUPPORTING INFORMATION

Examples of high-speed near-simultaneous nucleation images.